\def\rfr#1{eq. (\ref{#1})}

\def\eqi{\begin{equation}}
\def\eqf{\end{equation}}
\def\eqia{\begin{eqnarray}}
\def\eqfa{\end{eqnarray}}
\def\rp#1#2{{#1\over#2}} \def\lb#1{\label{#1}}

\documentclass[usenatbib]{mn2e}
\usepackage{amsmath}
\usepackage{amscd}
\usepackage{amssymb}
\usepackage{latexsym}
\usepackage{graphicx}
\usepackage{epsfig}

\title[Anthropic constraints on the cosmological constant from solar Galactic motion]{Anthropic constraints on the cosmological constant from Sun's motion through the Milky Way}

\author[L. Iorio]{
L. Iorio$^{1}$\thanks{E-mail:
lorenzo.iorio@libero.it}\\
$^{1}$INFN-Sezione di Pisa, Viale Unit$\grave{\rm a}$ di Italia 68, 70125, Bari (BA), Italy
}

\begin{document}

\date{Accepted 2009 December 14. Received 2009 November 30; in original form 2009 November 9}


\maketitle

\label{firstpage}

%
%
%
%
%
%
%
%
%
%
%
%
\begin{abstract}
We tentatively look at anthropic constraints on the Cosmological Constant (CC) $\Lambda$ at galactic scales by investigating its influence on the motion of the Sun throughout the Milky Way (MW) for $-4.5\leq t \leq 0$ Gyr. In particular, we look at the Galactocentric distance at which the Sun is displaced at the end of the numerical integration of its equations of motion modified in order to include the effect of $\Lambda$ as well. Values of it placing our star at its birth at more than 10 kpc from the Galactic center (GC) are to be considered implausible, according to the current views on the  Galactic Habitable Zone (GHZ) on the metallicity level needed for stars' formation. Also values yielding too close approaches to GC should be excluded because of the risks to life's evolution coming from too much nearby supernov{\ae} (SN) explosions and Gamma Ray Bursts (GRB). We investigate the impact on our results of the uncertainties on both the MW model's parameters and the Sun's initial conditions, in particular the Hubble parameter $H_0$ and the Local Standard Rest (LSR) speed $\Theta_0$ accurate at $2\%$ and $6.2\%$ level, respectively. While $H_0=70.1$ km s$^{-1}$ Mpc$^{-1}$, $\Theta_0=254$ km s$^{-1}$ and $\Lambda \leq 10^{-55}$ cm$^{-2}$ locates the place of birth of the Sun at $19.6$ kpc from GC, the same values for $H_0$ and $\Lambda$, and $\Theta_0^{\rm max}=270$ km s$^{-1}$, places it at the plausible Galactocentric distance of $8.5$ kpc. $\Lambda = 10^{-54}$ cm$^{-2}$ and $\Lambda = 10^{-53}$ cm$^{-2}$ place the Sun at $10.6$ kpc and $18.7$ kpc, respectively.
\end{abstract}



\section{Introduction}\label{intro}
The Cosmological Constant (CC) $\Lambda$ \citep{CC} is a physical quantity with the dimensions of an inverse area which was introduced for the first time by \citet{Ein} who modified the field equations of his General Theory of Relativity (GTR) to obtain cosmological static solutions. CC has recently been  used to explain the observed accelerated expansion of the Universe \citep{Riess,Perlm} in the most simple and economical way. As a result,  the so-called standard $\Lambda$CDM model\footnote{CDM stands for Cold Dark Matter.} \citep{Bah} arose. In this framework, CC has been interpreted as yielding a vacuum energy density. From the WMAP-BAO-SN\footnote{The Wilkinson Microwave Anisotropy Probe (WMAP) \citep{WMAP} is a spacecraft, located at the Lagrangian point L$_2$, which is currently measuring differences in the temperature of the Cosmic Microwave Background (CMB) radiation across the full sky since 2001. The Baryon Acoustic Oscillations (BAO) refer to an overdensity or clustering of baryonic matter at certain length scales due to acoustic waves which propagated in the early Universe \citep{BAO}.
SN refers to the Type Ia supernov{\ae} \citep{SN} which are a particular kind of deflagrating stars whose peculiar properties enabled to obtain the first evidence for the Universe's accelerated expansion \citep{Riess,Perlm}.} estimated parameter \citep{resu} \eqi\Omega_{\Lambda}\equiv\rp{\rho_{\Lambda}}{\rho_{\rm crit}}=0.726\pm 0.015,\eqf where\footnote{$c$ and $G$ are the speed of light in vacuum and the Newtonian gravitational constant, respectively.}
 \eqi \rho_{\Lambda}\equiv\rp{c^2\Lambda}{8\pi G},\ \rho_{\rm crit}\equiv\rp{3 H_0^2}{8\pi G}=(9.2\pm 0.3)\times 10^{-30}\ {\rm g\ cm}^{-3},\eqf in which \citep{resu} \eqi H_0=70.1\pm 1.3\ {\rm km\ s^{-1}\ Mpc^{-1}}=(2.27\pm 0.04)\times 10^{-18}\ {\rm s}^{-1}\lb{hubb}\eqf is the present-day value of the Hubble parameter, it turns out
\eqi \rho_{\Lambda}=(6.7\pm 0.3)\times 10^{-30}\ {\rm g\ cm}^{-3},\eqf and \eqi\Lambda =(1.26\pm 0.05)\times 10^{-56}\ {\rm cm}^{-2}.\lb{Lamb}\eqf

Such a small value is about 120 orders of magnitude smaller than the one theoretically computed, as vacuum energy density, in the framework of the currently accepted quantum field theories (QFT) \citep{Wein,Carroll}.

Even before the observational discovery of the cosmic acceleration
it was speculated on the constraints on a possible non-zero-and non necessarily positive-CC which could be posed by some physical properties of the Universe which allowed for the arising of life on the Earth.
For example, it was noted \citep{Bar,Lin} that if   $\rho_{\Lambda}\ll -10^{-29}$ g cm$^{-3}$, i.e. if CC was negative and quite small (that is, quite large in magnitude), then such a Universe,
even if  flat, would collapse within a time smaller than the age of our universe $t\sim 14$ Gyr, thus making our life, perhaps, impossible.
A more recent analysis \citep{Kal} points toward slightly less tight constraints: $\rho_{\Lambda}\gtrsim -2\times 10^{-28}$ g cm$^{-3}$ if 7 Gyr are sufficient for emergence of human life, while
$\rho_{\Lambda}\gtrsim -5\times 10^{-29}$ g cm$^{-3}$ if we really need 14 Gyr.
In fact, we may do not need 14 Gyr. Indeed, according to \citet{Line1}, Earth is a latecomer about 2 Gyr younger than an
average Earth-like planet in the Universe. Of course, we still have very confuse ideas about the timescales for the evolution of intelligent observers,
but there are indications that it could, in principle, occur significantly faster than it did on Earth (and in any case, it is reasonable to expect the timescales on individual planets to be broadly distributed).

Let us, now, examine anthropic constraints on positive CC.
It was initially argued
that the life of our type is impossible for $\rho_{\Lambda}\gg 10^{-29}$
g cm$^{-3}$ because, in this case, the density of matter of the
Universe would be exponentially small due to its exponential
expansion at the present stage \citep{Lin}.
 A tighter bound was obtained from the fact that $\Lambda$ must not be so high that galaxies never form \citep{Wein2} in the sense that if $\Lambda$ begins to dominate before the epoch of galaxy formation, the Universe will be devoid of galaxies, and thus of stars and planets. As a result, it should be $\rho_{\Lambda} \lesssim 10^{-27}$ g cm$^{-3}$. For anthropic constraints and time-varying fundamental constants, see \citet{Barro}.

 In this paper we want to try to preliminarily look at different anthropic constraints on $\Lambda$ by working at  non-cosmological scales. We will investigate the influence of CC on the motion of the Sun throughout the Galaxy backward in time, for $-4.5< t <0$ Gyr, to see where CC displaces our star at the epoch of its likely formation for given sets of initial conditions\footnote{An analogous approach has been recently applied by \citet{Iorio} to the past Galactic motion of the Sun according to different gravity models (without CC).}. Indeed, according to the concept of Galactic Habitable Zone (GHZ) \citep{Linea,Circo,Gonza,Bla,Pra}, the probability of having
Earth formed 4 and 8 Gyr after the formation of the Milky Way (MW),
which roughly corresponds to the birth of the Sun by assuming
a MW age of about 10-12 Gyr, is practically
null at Galactocentric distances larger than 10-15 kpc because of too poor metallicity \citep{Pra}. Thus, if a given value of $\Lambda$ displaced the Sun at a such huge distance, it should be regarded as implausible.
For different anthropic arguments applied to the Galaxy, see, e.g., \citep{Circo2}  and references therein.
Conversely, by taking the cosmologically/astrophysically-inferred values for CC and the MW's model parameters, our approach can also be used to put constraints on other features of the Galaxy like the present-day state vector of the Sun.

It must be stressed that our analysis, and the resulting conclusions, should be considered as preliminary because of the following considerations. First,
we do not include the Galactic model feedback for a chosen value of CC through dependence of the initial conditions  of the Galaxy formation.
Indeed, given the measured value of the total cosmological density, different values of CC would yield, in principle, different values of, say, the total present-day mass of the Galaxy, thus causing different present-day solar state vector and different GHZ boundaries at $t=-4.5$ Gyr. Second, the concept of GHZ itself lacks robustness since it has not been elucidated precisely enough so far, as admitted, among others, by \citet{Gonza} himself, one of the pioneers of the GHZ idea. It could also be argued that entrance/exit from GHZ may not be necessarily problematic. Indeed, \citet{Linea} frame GHZ as a part of the Galactic disk with a highest probability of forming solar-like planetary systems. If an already formed habitable system, i.e. of fixed chemical composition, leaves the GHZ, in particular beyond the outer boundary where far less hazardous threats should occur, it does not necessarily mean that it should become inhabitable. Another potential issue consists of the fact that the chemical evolution of the Galaxy is not, in principle, independent of CC. Indeed, different values of $\Lambda$ may shift the ratio of baryonic to non-baryonic matter (presumably CDM) in the original density perturbations, thus impacting the amount of baryons available for star formation, subsequent chemical enrichment, etc. However, in view of the lingering lacking of more detailed knowledge about such a topic, we will neglect it.

The outline of the paper is as follows. In Section \ref{due} we describe the model adopted for the Galaxy and the current state vector for the Sun which will constitute the initial conditions of our analysis. The results of our numerical integrations are presented in Section \ref{tre}, with a detailed discussion of the impact of the current uncertainties in the MW's model parameters and the Sun's initial conditions (Section \ref{trebis}). Section \ref{qua} is devoted to the discussion and the conclusions.
%
%
\section{Setting the scene}\label{due}
In order to properly describe the motion of the Sun through the Galaxy, we will adopt
the CDM model tested by \citet{Xue} with several  Blue Horizontal-Branch (BHB) halo stars. It  consists of  three components. Two of them are for the disk
\eqi U_{\rm disk}=-\rp{GM_{\rm disk}\left[1-\exp\left(-\rp{r}{b}\right)\right]}{r},\lb{disk}\eqf where $b$ is the disk scale length,
and the bulge
\eqi U_{\rm bulge} = -\rp{GM_{\rm bulge}}{r+c_0},\lb{bulge}\eqf
where $c_0$ is the bulge scale radius.
The third component is for the CDM NFW halo \citep{NFW}
\eqi U_{\rm NFW}=-\rp{4\pi G\rho_s r_{\rm vir}^3}{d^3 r}\ln\left(1+\rp{dr}{r_{\rm vir}}\right),\lb{halo}\eqf
with  $r_{\rm vir}$ is the radius parameter  and
\eqi \rho_s=\rp{\rho_{\rm crit}\Omega_{\rm m}\delta_{\rm th}}{3}\rp{d^3}{\ln\left(1+d\right)-\rp{d}{1+d}},\eqf
in which $\Omega_{\rm m}$ is the fraction of matter (including baryons and DM) to the critical density, $\delta_{\rm th}$ is critical overdensity of the virialized system, $d$ is the concentration parameter
%
%
%
%
The values used for the parameters entering \rfr{disk} and \rfr{bulge} are in Table \ref{buldis};
\begin{table}
\caption{Parameters of the disk and bulge models by \citet{Xue}.}
\label{buldis}
\begin{tabular}{cccc}\hline
$M_{\rm disk}$ & $M_{\rm bulge}$  & $b$  & $c_0$ \\
(M$_{\odot}$) & (M$_{\odot}$)  & (kpc)  & (kpc) \\
\hline
 $5\times 10^{10}$   & $1.5\times 10^{10}$  & $4$  & $0.6$\\
\hline
\end{tabular}
\end{table}
those entering \rfr{halo} are in Table \ref{NFW}.
\begin{table}
\caption{Parameters of the CDM NFW halo model by \citet{Xue}. The values quoted for $r_{\rm vir}$ and $d$ come from an average of those by \citet{Xue}.}
\label{NFW}
\begin{tabular}{ccccc}\hline
$\Omega_{\rm m}$ & $\delta_{\rm th}$  & $r_{\rm vir}$  & $d$ & $H_0$\\
- & -  & (kpc)  & (kpc) & (km s$^{-1}$ Mpc$^{-1}$)\\
\hline
 $0.3$   & $340$  & $273.2$  & $8.9$  & 65\\
\hline
\end{tabular}
\end{table}
Note that the value of the total baryonic mass of MW of Table \ref{buldis} is in agreement with the estimates by, e.g., \citet{McG} and \citet{Smi}.
In addition to the main acceleration coming from \rfr{disk}-\rfr{halo} we will add the contribution by CC. It is widely recognized that its effects are equivalent locally, i.e., within the distances of galaxies or galactic clusters, to those corresponding to a
repulsive tidal force, of a conservative nature, being derived from a unique scalar potential $U_{\Lambda}$ of the form \citep{Rin}
\eqi U_{\Lambda} = -\rp{c^2\Lambda}{6}r^2.\eqf
The initial conditions \citep{Reid} for our numerical integration backward in time of the Sun's equations of motion, performed with MATHEMATICA in rectangular Cartesian coordinates, are in listed in Table \ref{statevec}. For the speed of the Local Standard Rest (LSR) \citep{Binn}  we initially use the value $\Theta_0=220$ km s$^{-1}$ recommended by the International Astronomical Union\footnote{Decision taken by Commission 33 (Structure and Dynamics of the Galactic System/Structure et Dynamique du Syst\`{e}me Galctique) at the XIX General Assembly, Dehli, India, 1985. http://www.iau.org/static/resolutions/IAU1985$\_$French.pdf} (IAU).
\begin{table}
\caption{Galactocentric initial conditions for the Sun \citep{Reid}; the positive $y$ axis is directed from the Galactic Center (GC) to the Sun, the positive $x$ axis is directed toward the Galactic rotation, the positive $z$ axis is directed toward the  North Galactic Pole (NGP). We initially use the standard IAU value $\Theta_0 = 220$ km s$^{-1}$ for the rotation speed of LSR, and $U_0=10.3$ km s$^{-1}$, $V_0=15.3$ km s$^{-1}$, $W_0=7.7$ km s$^{-1}$ for the standard solar motions toward GC, $\ell=90$ deg and NGP, respectively; thus, with our conventions, $\dot x_0=V_0+\Theta_0,\ \dot y_0 = -U_0,\ \dot z_0 =W_0$. See Fig. 7 by \citet{Reid}.}
\label{statevec}
\begin{tabular}{ccccccc}\hline
$x_0$ & $y_0$  & $z_0$  & $\dot x_0$ & $\dot y_0$ & $\dot z_0$\\
(kpc) & (kpc)  & (kpc)  & (km s$^{-1}$) & (km s$^{-1}$) & (km s$^{-1}$) \\
\hline
 $0$  & $8.5$  & $0.02$  & $15.3+220$  & $-10.3$  & $7.7$  \\
\hline
\end{tabular}
\end{table}
\section{Results of the numerical integrations}\lb{tre}
Concerning negative values of CC, for $\Lambda=-10^{-50}$ cm$^{-2}$ it turns out the Sun would repeatedly pass as near as 2.5 kpc to GC,
although it would be finally located at 8.4 kpc at $t = -4.5$ Gyr. This would represent a serious danger for the birth of life on our planet because of the risk of a higher number of supernov{\ae} explosions and Gamma Ray Bursts (GRB); indeed, they may (partially or totally) destroy the Earth's atmospheric ozone,
leaving land life exposed to lethal does of ultraviolet (UV) fluxes from the Sun \citep{gamma}. Instead, for $\Lambda=-10^{-53}$ cm$^{-2}$ we have that 4.5 Gyr ago the Galactocentric distance of the Sun would amount to 14 kpc, outside the GHZ.  Larger values of $\Lambda$ (i.e. smaller values of its magnitude) would not pose problems because the resulting solar trajectories would substantially never pass below the 8 kpc level, displacing the Sun at about $8.5-8.8$ kpc at $t=-4.5$ Gyr. In all the cases examined departures from the Galactic plane would be of the order of $0.1-0.4$ kpc.

 Let us, now, move to positive values of CC, which is the most interesting situation in view of the present-day wealth of cosmological observations. It turns out that the value of \rfr{Lamb} for $\Lambda$ does not yield hazardous approaches to GC for $-4.5\leq t\leq 0$ Gyr and puts the Sun at $8.88$ kpc for $t=-4.5$ Gyr; for smaller values and for $\Lambda\rightarrow 0$ the Sun would be just at $8.88$ kpc as well. A CC one order of magnitude larger would slightly change the situation, displacing the Sun at $8.95$ kpc from GC. $\Lambda = 10^{-54}$ cm$^{-2}$ locates the Sun at $9.73$ kpc, while $\Lambda = 10^{-53}$ cm$^{-2}$ would be fatal since the Sun would be at  $15.8$ kpc. Also in this cases departures from the Galactic plane would be as large as about 0.5 kpc, and no closest approaches to GC would occur.
\subsubsection{The impact of the uncertainties in the MW's model and in the Sun's initial conditions}\lb{trebis}
It must be pointed out that our results depend in a non-negligible way both on the initial conditions of the Sun and on the MW model's parameters.

Indeed, if we adopt the best estimate for the latest kinematical determination\footnote{It has been obtained with Very Long Baseline Array (VLBA) and the Japanese VLBI Exploration of Radio Astronomy project measuring trigonometric parallaxes and proper motions of masers found in high-mass star-forming regions across MW.} of the LSR speed $\Theta_0=254$ km s$^{-1}$ by \citet{Reid}, we have that, for $\Lambda\rightarrow 0$, the final Sun's Galactocentric distance is 10.2 kpc, substantially the same value which can be obtained for values of $\Lambda$ equal to \rfr{Lamb} or smaller. $\Lambda=10^{-55}$ cm$^{-2}$ locates the Sun at $9.9$ kpc, while for $\Lambda=10^{-54}$ cm$^{-2}$  it would be at  $8.5$ kpc. $\Lambda=10^{-53}$ cm$^{-2}$ would, instead, displace the Sun at more than 20 kpc.

If we insert the value of \rfr{hubb} for the Hubble parameter in the MW model adopted, by keeping $\Theta_0=254$ km s$^{-1}$, we obtain that the Sun would be at 19.6 kpc for $\Lambda\rightarrow 0$ and \rfr{Lamb}. If $\Lambda$ was one order of magnitude larger, the Sun would be at 19.7 kpc. It would be at 20.5 kpc for $\Lambda=10^{-54}$ cm$^{-2}$. Curiously, $\Lambda=10^{-53}$ cm$^{-2}$, i.e. three orders of magnitude larger than \rfr{Lamb}, puts the Sun at 8.6 kpc.

Finally, the Hubble parameter of \rfr{hubb} and the standard IAU value $\Theta_0=220$ km s$^{-1}$ yield the Sun at $14.4$ kpc for $\Lambda\rightarrow 0$ and \rfr{Lamb}. A CC one-two orders of magnitude larger does not substantially alter the situation (14.4 kpc and 14.3 kpc, respectively), while also in this case $\Lambda=10^{-53}$ cm$^{-2}$ displaces the Sun at 8.6 kpc.

Let us, now, investigate the consequences of the uncertainty in the LSR speed amounting to $\delta \Theta_0= 16$ km s$^{-1}$ \citep{Reid}. By using \rfr{hubb} for the Hubble parameter and the upper limit on the LSR speed $\Theta_0^{\rm max}= 270$ km s$^{-1}$, it turns out that the final Sun's Galactocentric distance amounts to $8.5$ kpc for $\Lambda\rightarrow 0$. Also   $\Lambda \leq 10^{-55}$ cm$^{-2}$ yields the same value. Instead, $\Lambda = 10^{-54}$ cm$^{-2}$  places the Sun at 10.6 kpc, while it is found at 18.7 kpc for $\Lambda = 10^{-53}$ cm$^{-2}$. Concerning the lower limit $\Theta_0^{\rm min}= 238$ km s$^{-1}$, for $\Lambda\rightarrow 0$ and $\Lambda \leq 10^{-55}$ cm$^{-2}$ the Sun is at $16.7$ kpc. $\Lambda = 10^{-54}$ cm$^{-2}$ puts the Sun at $15.9$ kpc, while for $\Lambda = 10^{-53}$ cm$^{-2}$ it is at $11.9$ kpc.

If one uses the value by \citet{Xue} for $H_0$, the results are quite different. $\Theta_0^{\rm max}$ and $\Lambda\rightarrow 0$ displace the Sun at $27.7$ kpc. The same holds for $\Lambda \leq 10^{-55}$ cm$^{-2}$. A reduction of about 1 kpc (26.4 kpc) occurs for $\Lambda = 10^{-54}$ cm$^{-2}$, while $\Lambda = 10^{-53}$ cm$^{-2}$ locates the Sun at $29.7$ kpc.
For $\Theta_0^{\rm min}$ and $\Lambda\rightarrow 0$, the Sun is at $13.1$ kpc, as for $\Lambda \leq 10^{-55}$ cm$^{-2}$. The solar Galactocentric distance rises to $15.1$ kpc for  $\Lambda = 10^{-54}$ cm$^{-2}$, while $\Lambda = 10^{-53}$ cm$^{-2}$ it is $13.9$ kpc.
\section{Summary and conclusions}\label{qua}
We have preliminarily investigated the possibility of obtaining anthropic constraints on a constant and uniform CC at a galactic scale. Conversely, by keeping the cosmologically determined values of $\Lambda$ and $H_0$, we can use our analysis to constrain some properties of the galactic system considered.

 We considered the Sun under the action of the gravitational field of MW, modeled in terms of bulge+disk+NFW CDM halo, with the addition of the action of such kind of CC. We numerically integrated the resulting equations of motion backward in time from now to 4.5 Gyr ago. We looked at the final Galactocentric distance of the Sun at the epoch of its likely formation in view of the limitations on the Earth's formation and evolution of complex forms of life on it  outlined within the GHZ framework.

Concerning negative values for CC, the anthropic constraints that can be derived are much tighter than those previously obtained in literature. However, we focussed most of our analysis on positive CC.

 Our results are, in general, strongly influenced by the current uncertainties in both the adopted MW model's parameters and Sun's initial conditions, in particular by the Hubble parameter $H_0$ and by the LSR speed $\Theta_0$. Table \ref{resume} resumes the numerical outcomes of our analysis.
\begin{table}
\caption{Final Galactocentric Sun's distance $R$, in kpc, at $t=-4.5$ Gyr for different values of the solar, Galactic and cosmological relevant parameters. Positive values of CC are considered.}
\label{resume}
\begin{tabular}{ccccc}\hline
$H_0$ (km s$^{-1}$ Mpc$^{-1}$) & $\Lambda$  (cm$^{-2}$) & $\Theta_0$ (km s$^{-1}$) &  $R$ (kpc)\\
\hline
$65$ & $\leq 10^{-56}$ & $220$ & $8.88$  \\
$65$ & $10^{-55}$ & $220$ & $8.95$  \\
$65$ & $10^{-54}$ & $220$ & $9.73$  \\
$65$ & $10^{-53}$ & $220$ & $15.8$  \\
\hline
$65$ & $\leq 10^{-56}$ & $254$ & $10.2$  \\
$65$ & $10^{-55}$ & $254$ & $9.9$  \\
$65$ & $10^{-54}$ & $254$ & $8.5$  \\
$65$ & $10^{-53}$ & $254$ & $>20$  \\
\hline
$70.1$ & $\leq 10^{-55}$ & $220$ & $14.4$  \\
$70.1$ & $10^{-54}$ & $220$ & $14.3$  \\
$70.1$ & $10^{-53}$ & $220$ & $8.6$  \\
\hline
$70.1$ & $\leq 10^{-56}$ & $254$ & $19.6$  \\
$70.1$ & $10^{-55}$ & $254$ & $19.7$  \\
$70.1$ & $10^{-54}$ & $254$ & $20.5$  \\
$70.1$ & $10^{-53}$ & $254$ & $8.6$  \\
\hline
$65$ & $\leq 10^{-55}$ & $270$ & $27.7$  \\
$65$ & $10^{-54}$ & $270$ & $26.4$  \\
$65$ & $10^{-53}$ & $270$ & $29.7$  \\
\hline
$65$ & $\leq 10^{-55}$ & $238$ & $13.1$  \\
$65$ & $10^{-54}$ & $238$ & $15.1$  \\
$65$ & $10^{-53}$ & $238$ & $13.9$  \\
\hline
$70.1$ & $\leq 10^{-55}$ & $270$ & $8.5$  \\
$70.1$ & $10^{-54}$ & $270$ & $10.6$  \\
$70.1$ & $10^{-53}$ & $270$ & $18.7$  \\
\hline
$70.1$ & $\leq 10^{-55}$ & $238$ & $16.7$  \\
$70.1$ & $10^{-54}$ & $238$ & $15.9$  \\
$70.1$ & $10^{-53}$ & $238$ & $11.9$  \\
\hline
\end{tabular}
\end{table}

 For all the values used for them it turns out that the $\Lambda$CDM value of $\Lambda$ yields solar Galactocentric final distances which are indistinguishable  from the CC$=0$ case.  $H_0=65$ km s$^{-1}$ Mpc$^{-1}$ and $\Theta_0=220$ km s$^{-1}$ locate the Sun at 8.88 kpc, while it is at 10.2 kpc for $\Theta_0=254$ km s$^{-1}$.  Instead, $H_0=70.1$ km s$^{-1}$ Mpc$^{-1}$ displaces the Sun at 19.6 kpc for  $\Theta_0=254$ km s$^{-1}$ and at 14.4 kpc for $\Theta_0=220$ km s$^{-1}$. In the first case the Sun's Galactocentric distance is satisfactory, but it is based on values of $H_0$ and $\Theta_0$ which are in contrast with
 the latest observational determinations. The third case relies upon them, but the resulting place of birth of the Sun is too far from GC; anyway, it must be pointed out that the latest determinations of $\Theta_0$ \citep{Reid} are uncertain at more than $6\%$ level, and this can make the difference.

 In general, significative variations in the Sun's Galactocentric distances with respect to the standard $\Lambda$CDM  case ($\Lambda = 10^{-56}$ cm$^{-2}$) occur for values of CC larger by two-three orders of magnitude. In particular, we note that the recently determined values $H_0=70.1$ km s$^{-1}$ Mpc$^{-1}$, $\Theta_0=254$ km s$^{-1}$ and a CC three orders of magnitude larger than the $\Lambda$CDM value, i.e. $\Lambda = 10^{-53}$ cm$^{-2}$, put the Sun at 8.6 kpc. The same occurs also for $\Theta_0=220$ km s$^{-1}$, given the values of the other parameters unchanged, i.e. $H_0=70.1$ km s$^{-1}$ Mpc$^{-1}$ and $\Lambda = 10^{-53}$ cm$^{-2}$. Anyway, such results, obtained for the best estimates of $\Theta_0$, which is actually uncertain at a $6.2\%$ level, are in disagreement with the $\Lambda$CDM cosmological value for $\Lambda$ by three orders of magnitude.

 A final Galactocentric distance of 8.5 kpc is also given by\footnote{The value $\Theta_0^{\rm max}=270$ km s$^{-1}$ by \citet{Reid} is close to $V_{\rm slr}=273.9$ km s$^{-1}$ obtained by \citet{Liu} from the proper motion and parallax data for 1011 O-B stars in the Hipparcos Catalogue.} $H_0=70.1$ km s$^{-1}$ Mpc$^{-1}$, $\Theta_0^{\rm max}=270$ km s$^{-1}$ and  $\Lambda\leq 10^{-55}$ cm$^{-2}$ which includes the $\Lambda$CDM value as well. Instead, the smallest value for the LSR speed $\Theta_0^{\rm min}=238$ km s$^{-1}$ is disfavored since it places the Sun at 16.7 kpc for $H_0=70.1$ km s$^{-1}$ Mpc$^{-1}$ and $\Lambda\leq 10^{-55}$ cm$^{-2}$. Thus, if we accept the WMAP-BAO-SN values of $H_0$ and $\Lambda$, accurate at a $2-4\%$ level, we can use our analysis to put tighter constraints on the Galactic LSR speed which would be, thus, closer to $\Theta_0^{\rm max}$ than to $\Theta_0^{\rm min}$. Generally speaking, it would be interesting to process again the Galactic observations with dynamical models modified by the introduction of a CC, but it is beyond the scopes of the present work. Conversely, if we take the current value for $H_0$ and $\Theta_0^{\rm max}$, it turns out that a CC one-two orders of magnitude larger than its standard $\Lambda$CDM value remains acceptable, yielding a reasonable solar Galactocentric distance at $t=-4.5$. Gyr.

In conclusion, given all the unavoidable limitations of the approach followed here outlined in the Introduction, we can say that for $H_0=65$ km s$^{-1}$ Mpc$^{-1}$ and $\Theta_0=220/254$ km s$^{-1}$, favorable anthropic Galactocentric distances $R\lesssim 10$ kpc are obtained for $\Lambda\leq 10^{-54}$ cm$^{-2}$. For $H_0=70.1$ km s$^{-1}$ Mpc$^{-1}$ the situation is more intricated because of the impact of the uncertainties in $\Theta_0$. Indeed, we have that for $\Theta_0=220/254$ km s$^{-1}$, $\Lambda=10^{-53}$ cm$^{-2}$ yields $R< 10$ kpc, while for $\Theta_0=270$ km s$^{-1}$ it must be $\Lambda\leq 10^{-55}$ cm$^{-2}$.
\section*{Acknowledgments}
I thank an anonymous referee for constructive remarks. I am also grateful to F. Scardigli and M. Della Valle for interesting and useful discussions. The support of ASI is acknowledged.


\end{document}